\documentclass[a4paper,11pt]{article}
\usepackage{arxiv}
\pdfoutput=1
\usepackage[utf8]{inputenc} 
\usepackage[T1]{fontenc} 
\usepackage{geometry}
\usepackage{setspace}
\usepackage{authblk}
\usepackage{hyperref} 
\usepackage{url} 
\usepackage{graphicx}
\usepackage[shortlabels]{enumitem}
\usepackage{caption}
\usepackage[scaled=1]{helvet}

\usepackage{pdflscape}
\usepackage{longtable} 
\usepackage{booktabs} 
\usepackage{lscape} 
\usepackage{array} 
\usepackage{multirow}
\usepackage{colortbl}
\usepackage{xcolor}
\usepackage{amsfonts} 
\usepackage{dsfont} 
\usepackage{microtype} 
\usepackage{lipsum} 
\usepackage{doi}
\usepackage{framed}
\usepackage{longtable}
\usepackage{pdfpages}
\usepackage{amsmath,amssymb}
\usepackage{calc}
\usepackage{mathtools}
\graphicspath{{/home/marc/Documents/RPROJ_2/CAD_stroke/reports/TEX/ARXIV/PICS/}}

\author[1,*]{Marc Delord}
\author[1]{Xiaohui Sun}
\author[1]{Annastazia Learoyd}
\author[1]{Vasa Curcin}
\author[1]{Iain Marshall}
\author[1]{\\Charles Wolfe}
\author[1]{Mark Ashworth}
\author[1]{Abdel Douiri}

\affil[1]{School of Life Course \& Population Sciences, Department of Population Health Sciences, King's College London, London, United Kingdom}
\affil[*]{Correspondence: \href{mailto:marc.delord@kcl.ac.uk}{marc.delord@kcl.ac.uk}}

\title{Patient-Oriented Unsupervised Learning\\ to Unlock Patterns of Multimorbidity Associated with Stroke\\ using Primary Care Electronic Health Records}

\hypersetup{
pdftitle={Unsupervised Learning using EHRs to Unlock Patterns Multimorbidity Associated with Stroke},
pdfsubject={stat.AP, stat.ME, stat.ML},
pdfauthor={Marc Delord},
pdfkeywords={Electronic Health Records, Multimorbidity, Stroke, Multiple State Analysis, Jaccard Dissimilarity Index , Ward clustering},
}
\renewcommand{\shorttitle}{\scriptsize Unsupervised Learning using EHRs to Unlock Patterns Multimorbidity Associated with Stroke}

\date{}

\begin{document}
\maketitle
\setstretch{1.3}
\begin{abstract}
\noindent
\textbf{Background}: Identifying and characterising the longitudinal patterns of multimorbidity associated with stroke is needed to better understand patients’ needs and inform new models of care.

\noindent
\textbf{Methods}: We used an unsupervised patient-oriented clustering approach to analyse primary care electronic health records (EHR) of 30 common long-term conditions (LTC), in patients with stroke aged over 18, registered in 41 general practices in south London between 2005 and 2021.

\noindent
\textbf{Results}: Of 849,968 registered patients, 9,847 (1.16\%) had a record of stroke, 46.5\% were female and median age at record was 65.0 year (IQR: 51.5 to 77.0). The median number of LTCs in addition to stroke was 3 (IQR: from 2 to 5). Patients were stratified in eight clusters. These clusters revealed contrasted patterns of multimorbidity, socio-demographic characteristics (age, gender and ethnicity) and risk factors. Beside a core of 3 clusters associated with conventional stroke risk-factors, minor clusters exhibited less common but recurrent combinations of LTCs including mental health conditions, asthma, osteoarthritis and sickle cell anaemia. Importantly, complex profiles combining mental health conditions, infectious diseases and substance dependency emerged.

\noindent
\textbf{Conclusion}: This patient-oriented approach to EHRs uncovers the heterogeneity of profiles of multimorbidity and socio-demographic characteristics associated with stroke. It highlights the importance of conventional stroke risk factors as well as the association of mental health conditions in complex profiles of multimorbidity displayed in a significant proportion of patients. These results address the need for a better understanding of stroke-associated multimorbidity and complexity to inform more efficient and patient-oriented healthcare models.
\end{abstract}
\break

\setstretch{1.3}
\section{Introduction}

Despite recent attention to the topic, current knowledge of multimorbidity in stroke remains limited \cite{gallacher2014stroke,Nelson2015,gallacher2018risk,gallacher2019multimorbidity}. Multimorbidity is commonly defined as the co-occurrence of two or more long-term conditions (LTCs) \cite{johnston2019defining,skou2022multimorbidity}. This non-specific definition encompasses conventional stroke risk factors such as hypertension, diabetes and non-conventional stroke risk factors such as chronic inflammatory diseases or chronic mental health disorders, as well as LTCs less specifically associated with stroke such as osteoarthritis or asthma \cite{gallacher2019multimorbidity}. In the context of the rising prevalence of LTCs within an ageing population \cite{uijen2008multimorbidity,fortin2012systematic}, most patients with stroke present at least one or more additional LTC \cite{gallacher2014stroke,gallacher2019multimorbidity}. These factors challenge primary healthcare services, as multimorbidity increases patients complexity which may result in poorer short- and long-term outcome after stroke \cite{schmidt2014eighteen,gallacher2018risk} and increased costs of care \cite{gallacher2019multimorbidity,skou2022multimorbidity,safford2007patient}.

To date, the analyses of stroke related multimorbidity has been based on classical epidemiological approaches such as case-control studies focusing on the association of potential stroke risk factors with stroke status \cite{o2010risk,gallacher2014stroke,bang2015nontraditional,gallacher2019multimorbidity} or cross-sectional studies assessing the impact of different indices of multimorbidity, such as the Charlson Comorbidity Index scores \cite{schmidt2014eighteen} or the simple count of LTCs on stroke outcomes \cite{gallacher2014stroke,gruneir2016increasing, liu1997comorbidity,johnston2019defining}.

These approaches however remain general and beyond the necessary characterisation of risk factors and long-term conditions associated with stroke\cite{gallacher2014stroke,schmidt2014eighteen,gallacher2018risk,kelly2021impact}, the need arise for a better understanding of multimorbidity associated with stroke. This includes patients stratification according to the main patterns of multimorbidity, the longitudinal description of defined clusters, including their relative size, associated socio-demographic characteristics and risk factors \cite{gallacher2019multimorbidity,aquino2019does}.

The life course setting represents an alternative framework to meet this objective \cite{elder1998life} using primary care electronic health records (EHR) and unsupervised learning methods \cite{kaufman2009finding}. Unlike classical approaches used to analyse multimorbidity in cross-sectional studies \cite{ng2018patterns,sukumaran2023understanding} which result in non-specific descriptions of common LTCs associated with a given cohort of patients, these approaches allow to use longitudinal EHRs to define individual patient health trajectories using available patient health history \cite{pollock2007holistic}. Resulting patient individual health trajectories are further analysed as input in an unsupervised clustering procedure to identify sub-populations of patients (clusters) characterised by distinct life course health trajectories. In turn, the longitudinal description of these health trajectories in terms of longitudinal patterns of multimorbidity allows to better apprehend patients complexity addressing evidence-gaps toward patient-oriented approach of care delivery \cite{shippee2012cumulative,gallacher2019multimorbidity,aquino2019does,safford2007patient}.

In this study, we analysed longitudinal primary care EHRs related to up to 30 LTCs, including stroke, as well as major conventional stroke risk factors such as hypertension and diabetes, in 9,847 patients with stoke using a patient-oriented unsupervised learning approach. Our objective was to identify and characterise the main patterns of multimorbidity associated with stroke.


\section{ Results }

\subsection{ Patients characteristics and Long-term conditions }

Of 849,968 registered patients, 9,847 (1.16\%) had a record of stroke. This corresponds to an incidence rate of 166.37 cases per 100 000 person-years. The median age at record of stroke was 65.0 years. Among these patients, 46.5\% were female, 50.8\% were from White ethnicity and 40.9\% belonged to index of multiple deprivation (IMD) quintile 1-2 (most deprived) (vs quintile 3-5 (less deprived)).

\begin{figure}[!ht]
\centering
\makebox{\includegraphics[scale=.5, trim = 0cm 0cm 0cm 0cm, clip]{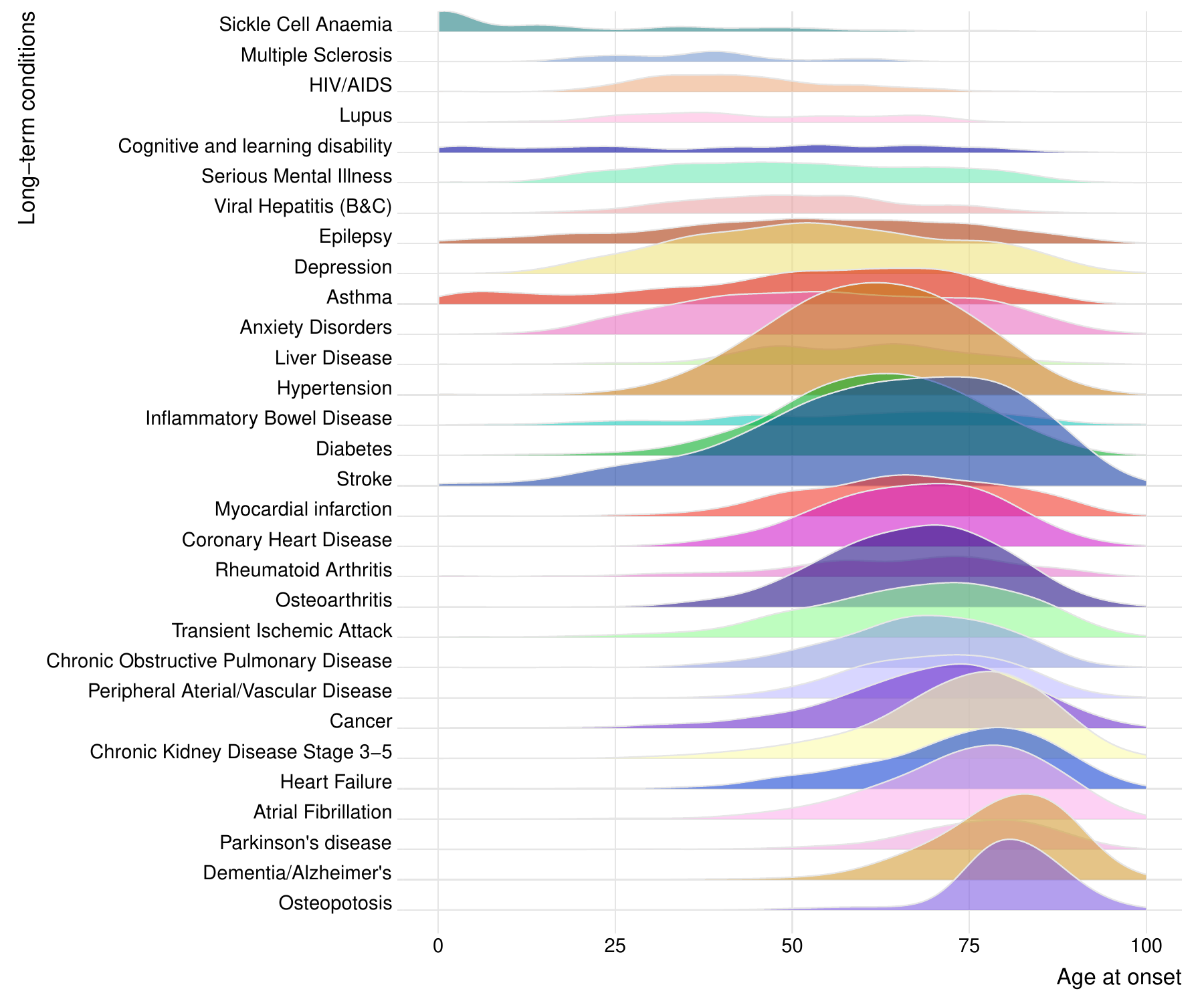}}
\caption{Scaled density of age at record of the 30 analysed long-term conditions. Long-term conditions are ordered by ascending median age at time of records from top to bottom.}
\label{fig:1}
\end{figure}

Along with hypertension (64.4\%) and diabetes (30.5\%), the other frequent LTCs were: osteoarthritis (26.2\%), depression (23.8\%), chronic kidney disease (23.3\%), anxiety (20.2\%), atrial fibrillation (17.0\%), coronary heart disease (16.8\%), transient ischemic attack (15.8\%) and cancer (15.6\%) (table \ref{tab:t1}). Figure \ref{fig:1} displays the scaled densities of patients' age at record of analysed LTCs.

\begin{figure}[!ht]
\centering
\makebox{\includegraphics[scale=.7, trim = 0cm 0cm 0cm 0cm, clip]{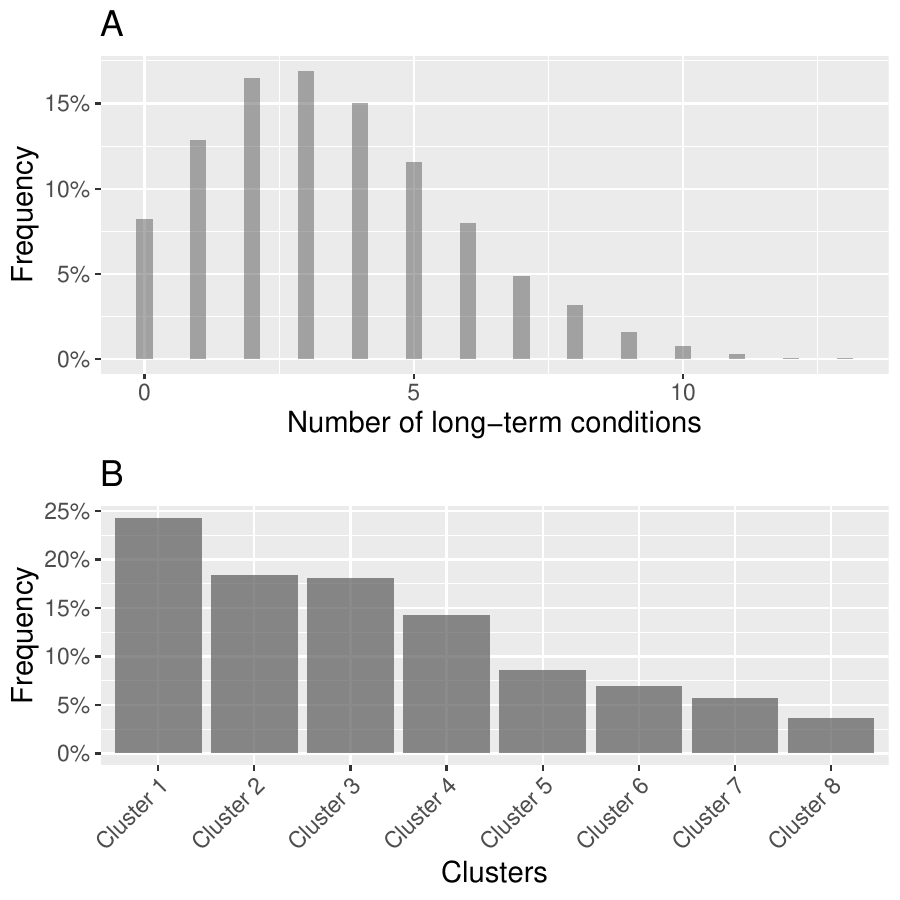}}
\caption{Distribution of the number of LTCs in addition to stroke in analysed patients (A), and size of identified clusters (B).}
\label{fig:2}
\end{figure}

In addition to stroke, the median number of LTCs was 3 (interquartile range from 2 to 5, figure \ref{fig:2}-A). The median age at end of follow-up was 50.1 years, 62.4\% of patients were censored at the end of follow-up whereas 37.6\% died (table  \ref{tab:t2}).

\subsection{ Clustering }

After computation of pairwise dissimilarities among patients and hierarchical clustering, the point-biserial correlation coefficient were computed for a set of partition sizes, ranging from two to 20. An eight-clusters partition was associated with a local maximum. Accordingly, clusters were ordered by decreasing frequency and numbered from 1 to 8 (Figure \ref{fig:2}-B, Tables \ref{tab:t1} and \ref{tab:t2}).

\cleardoublepage
\subsection{ Clusters characteristics and graphical representations }

Table \ref{tab:t2} displays the distribution of socio-demographic characteristics and risk factors across clusters, and table \ref{tab:t1} presents the distribution of LTCs across clusters. In analogy with figure \ref{fig:1}, supplementary figure S3 displays the scaled densities of patients' age at record of analysed LTCs across clusters.

\begin{figure}[!ht]
\centering
\includegraphics[trim = 0cm 0cm 0cm 0cm , clip,scale= .9]{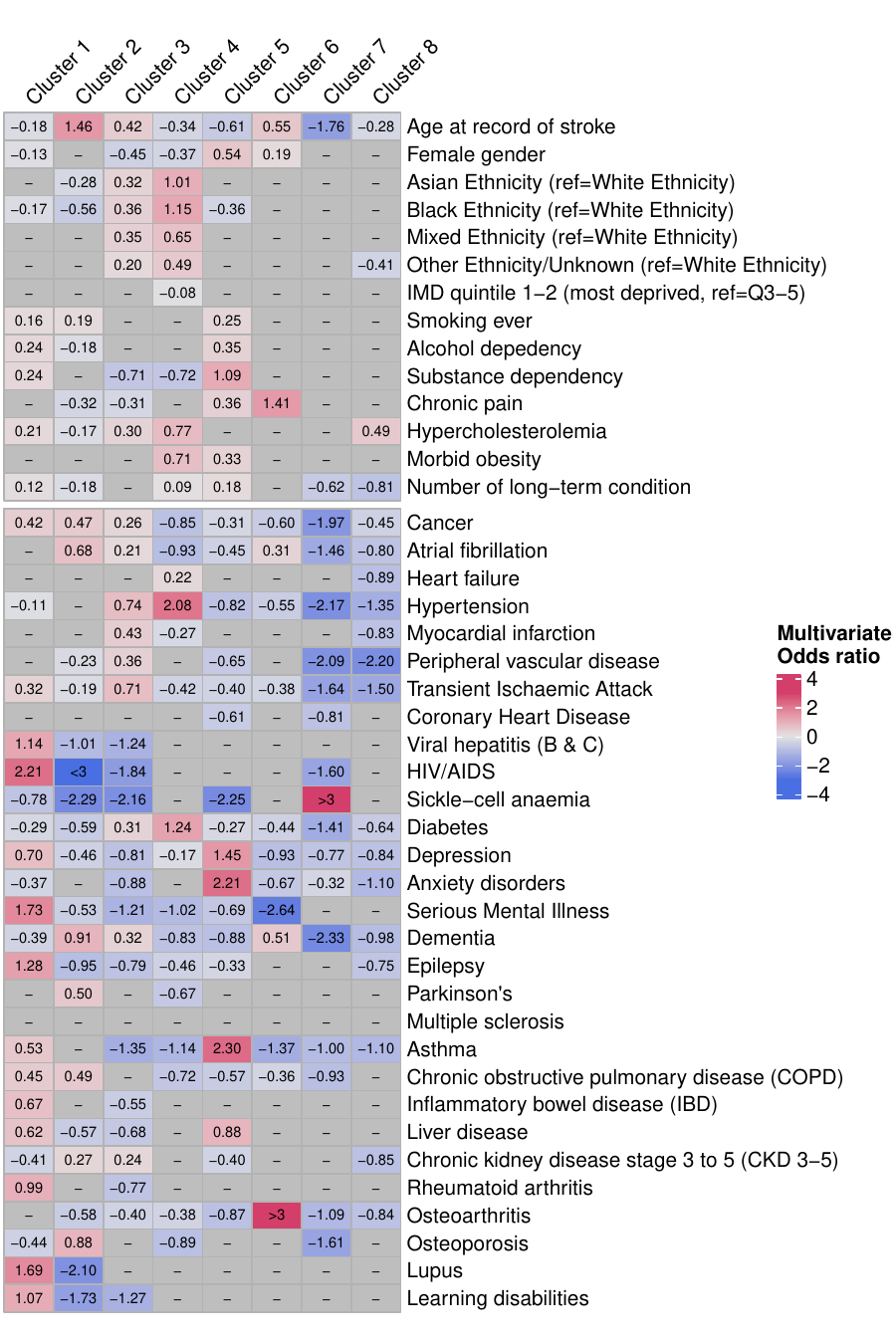}
\caption{Heatmap of log-odds ratio of socio-demographic variables and conventional stroke risk factors (upper panel) and LTCs (lower panel) associated with defined clusters: in both panels, values are derived from multivariate logistic regressions where cluster indicators are explained by displayed variables. Positive log-odds ratios (red versus blue) indicate over representation of the corresponding population traits in a given cluster as compared to its conditional expectation. Only coefficients associated with a P-value below the 5\% level were displayed.}
\label{fig:3}
\end{figure}

Figure \ref{fig:3} displays the multivariate log-odds ratios of socio-demographic characteristics and risk factors (upper panel) and  LTCs (lower panel) for each clusters relatively to all other stroke patients. This representation provides an overall representation of patients stratification. Finally, supplementary figure S4 represents clusters backbone by displaying the main sequences of LTCs associated with stroke across clusters.


\subsubsection{ Clusters examination }

From Tables \ref{tab:t2} and \ref{tab:t1}, figure 3 and supplementary figure S4, clusters can be characterised as follow:
\begin{itemize}

\item Cluster 1 (n=2392, 24.3\%) is the major cluster of the typology, comprising almost a quarter
 of patients. This cluster displays average socio-demographic characteristics and high prevalence of substances dependency (including smoking and alcohol), hypercholesterolaemia and LTCs in addition to stroke (median: 4), including cancer, infectious and inflammatory diseases, mental health conditions, and diseases of the respiratory system.

\item Cluster 2 (n=1814, 18.4\%) is characterised by the oldest median age at record of stroke (81.2 years) and under-representation of Asian and Black ethnicity. In contrast to cluster 1, this cluster exhibits a reduced prevalence of alcohol, substance dependency and hypercholesterolaemia while displaying LTCs characteristic of advanced age, such as cancer, atrial fibrillation, dementia, Parkinson's, and osteoporosis. Additionally, COPD (chronic obstructive pulmonary disease) and CKD 3-5 (chronic kidney disease stage 3 to 5) are also prevalent in this cluster, whereas infectious and inflammatory diseases, mental health conditions, liver disease, and osteoarthritis are less represented.

\item Cluster 3 (n=1782, 18.1\%) is defined by an older median age at record of stroke (70.7 years), non-White ethnicity, and a higher proportion of male patients. This cluster presents also a high prevalence of hypercholesterolaemia. LTCs associated with this cluster include cancer, cardiovascular conditions, diabetes, dementia, and CKD 3-5. Similarly to cluster 2, cluster 3 has lower prevalence of infectious and inflammatory diseases, mental health conditions, liver disease, and osteoarthritis.

\item Clusters 4 (n=1409, 14.3\%) and 3 share similar socio-demographic characteristics. However, records of stroke occur earlier in cluster 4, at the median age of 55 years, compared to 70.7 years in cluster 5. Cluster 4 presents also the highest prevalence of hypertension (92.3\%) and diabetes (59.0\%) and a higher proportion of patients from Asian, Black and mixed ethnicity.

\item Cluster 5 (n=845, 8.6\%) is characterized by a younger median age at record of stroke (47 years), with an over-representation of female patients and patients from White ethnicity. It also has higher prevalence of substances dependency (including smoking and alcohol) and chronic pain. Principal LTCs present in this cluster include depression, anxiety disorders, asthma, and liver disease.

\item Cluster 6 (n=685, 6.7\%) displays the second oldest median age at record of stroke (78.1 years) a slightly over representation of female patients. This cluster is also characterized by the highest prevalence of chronic pain and  LTCs associated with older age, such as atrial fibrillation and dementia. Nearly all patients in this cluster were diagnosed with osteoarthritis.

\item Cluster 7 (n=563, 5.7\%) displays the youngest median age at record of stroke (26.8 years), average socio-demographic characteristics, and the lowest median number of LTCs. The majority of patients diagnosed with sickle cell anaemia belong to this cluster.

\item Cluster 8 (n=357, 3.6\%) presents similar socio-demographic characteristic characteristics to cluster 7, but with a later median age at record of stroke (44.4 years) and a high prevalence of hypercholesterolaemia.

\end{itemize}

In terms of number of LTCs, clusters 1 and 5 present the highest level of multimorbidity with a median of 4 recorded LTCs, clusters 2 to 5 presents intermediate levels of multimorbidity with a median of 3 recorded LTCs, whereas clusters 7 and 8 present low profiles of multimorbidity, their median number of recorded LTC being zero (i.e., apart from stroke, no other LTCs were recorded for at least half of patients in these clusters) (table \ref{tab:t2}).

\section{ Discussion }

This study has adopted a unsupervised patient-oriented clustering approach to analyse primary care electronic health records (EHR) in patients with stroke to disentangle the patterns of multimorbidity associated with stroke. 30 LTCs (including stroke), selected according to a consensual definition of multimorbidity, were considered in a relatively large cohort of 9,847 patients with stroke. Accordingly, patients were stratified in eight clusters of health trajectories characterised by specific patterns of multimorbidity, socio-demographic characteristics, and risk factors.\\

\subsection{ Main findings }

Cluster 1 which represents about a quarter of analysed patients displays socio-demographic characteristics broadly representative of the whole sample, a higher prevalence of risk factors such as substances dependency and hypercholesterolaemia, and a high number of recorded LTCs including cancer, infectious/viral diseases, inflammatory diseases, respiratory system diseases and importantly mental health conditions.

\begin{figure}[!ht]
\centering
\includegraphics[trim = 0cm 0cm 0cm 0cm , clip,scale= .9]{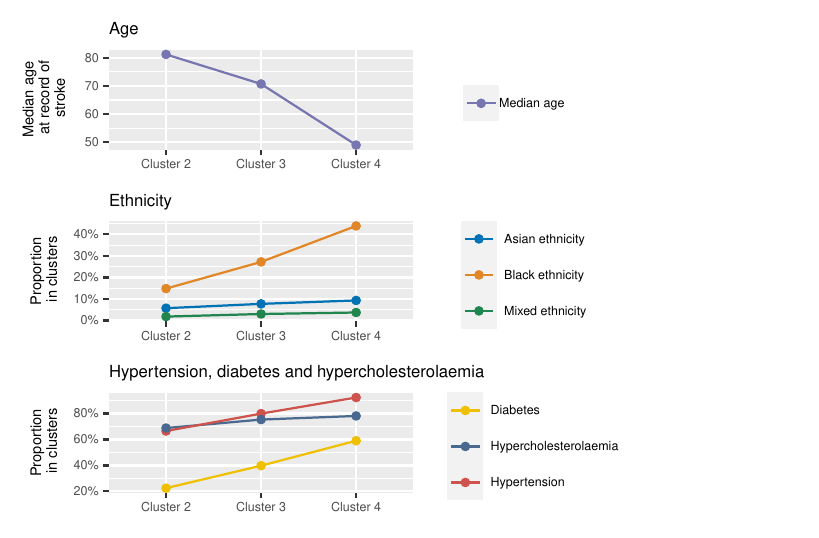}
\caption{Trends in the median age at record of stroke, the proportion of patients from Asian, Black and mixed ethnicity, and the prevalence of hypertension, diabetes and hypercholesterolaemia in clusters 2, 3 and 4.}
\label{fig:4}
\end{figure}

Clusters 2 to 4 represent more than half of analysed patients. On average, these clusters display expected characteristics such as older age, male gender and history of hypercholesterolaemia, hypertension and diabetes \cite{feigin2017global,richards2002antecedent}. While the age at record of stroke displays a downward trend from cluster 2 to 4, the prevalence of hypercholesterolaemia, hypertension, and diabetes, as well as the representation of Asian, Black and mixed ethnicity, follows an opposite trend (figure \ref{fig:4}).

Clusters 5 to 8 are minor clusters of the typology, representing approximately a quarter of analysed patients. These clusters exhibit distinct patterns in terms of socio-demographic characteristics, risk factors, and multimorbidity. Cluster 5 is characterised by younger age, female gender, White ethnicity, substance dependency, chronic pain, asthma, and mental health conditions. Cluster 6 features older age, female gender, chronic pain, and osteoarthritis. Cluster 7 and 8 are characterised by younger age, lower prevalence of conventional stroke risk factors and multimorbidity. Sickle cell anaemia is the sole LTCs specifically associated with cluster 7 and  hypercholesterolaemia is the only stroke risk factor associated with cluster 8.

Higher and intermediate levels of multimorbidity were encountered in clusters 1 and 6, and clusters 2 to 5, the median number of LTCs being 4 and 3 in these clusters respectively. On the other hand, clusters 7 and 8 (which represent up to 9.3\% of analysed patients) exhibit low multimorbidity profiles.

\subsection{Multimorbidity and patient complexity (clusters 1, 5 and 6)}

Higher levels of multimorbidity are known to have important implications for the design of health services \cite{cassell2018epidemiology} while increasing patients complexity \cite{safford2007patient,gallacher2019multimorbidity}. Patient complexity has been defined as a self-reinforced dynamic state in which the personal, social and clinical aspects of a patient experience interact \cite{shippee2012cumulative,may2014rethinking}. Multimorbidity impacts patient complexity in various ways. Examples includes the direct impact of LTCs on patient functional capacity (e.g., the limitation in usual activities due to osteoarthritis symptoms \cite{centers2006prevalence}), interactions between LTCs/risk-factors \cite{moussavi2007depression}, treatment interactions \cite{guthrie2015rising} or treatment burden \cite{gallacher2018conceptual}. Importantly patient self-care and ability to cope with complexity, referred to as patient capacity, depends on factors including for instance physical/mental functioning, social support, literacy and attitude\cite{shippee2012cumulative,may2014rethinking}. If multimorbidity directly increases patient workload (also referred to as burden of treatment), it may also differentially affect and erode patient capacity according to specific patterns of multimorbidity (burden of illness). In this perspective, the qualitative and longitudinal description of stroke-associated multimorbidity reported in this study allows for a better understanding of patient complexity and its consequence on patient care \cite{shippee2012cumulative}.

For instance, LTCs such as diabetes, which require considerable effort on behalf of a patient to modify diet, engage in physical activity and monitor blood sugar \cite{russell2005time}, present high burden of treatment profiles. Conversely, mental health disorders present more complex profiles, combining high burden of treatment and high burden of illness. This can significantly affect patient capacity, through functional morbidity \cite{burdick2009attention,wagner2000minor} and poor adherence to treatment and medical appointments \cite{shippee2012cumulative,katon2005impact}.

Accordingly, clusters 2 to 4 which display higher levels of (manageable) conventional stroke risk factors can be seen as less complex than clusters 1, 5 and 6 which are enriched with LTCs displaying high burden of illness such as mental health disorders (clusters 1 and 5), learning disabilities (cluster 1) and osteoarthritis (cluster 6).

In addition to the description of clusters of multimorbidity, socio-demographic characteristics and risk-factors associated with clusters enable further characterisation of these clusters in terms of potential factors of complexity. Notably, clusters 1 and 5 display higher prevalence of substance dependency including tobacco and alcohol. These characteristics can be viewed not only as direct stroke risk factors \cite{de2012trends,o2016global} but also as additional factors of complexity, given that these behavioural/lifestyle characteristics represent also major risk factors to various other LTCs \cite{mertens2005hazardous}, including traumatic injuries \cite{Cherpitel2007,cheng2016impact}, HIV/AIDS \cite{anand2010neurocognitive}, hepatitis C \cite{loftis2006psychiatric}, mental health conditions \cite{kelly2013integrated} and cardiovascular diseases \cite{gan2021risk}.

In turn, these behavioural/lifestyle aspects sheds further light on the specific patterns of multimorbidity observed in associated clusters, notably the higher prevalence of HIV/AIDS and viral hepatitis as well as serious mental illness in cluster 1, and depression and anxiety disorder in cluster 5. Likewise, the higher prevalence of chronic pain observed in cluster 5 and 6 can be apprehended as a factor of complexity associated with depression, anxiety disorders and morbid obesity \cite{peppin2015complexity} in cluster 5 and osteoarthritis \cite{schaible2012mechanisms} in cluster 6.

As a consequence, clusters 1, 5 and 6 can be regarded as complex clusters. Yet, the combination of mental health conditions, substances dependency, and infectious diseases in cluster 1 adds an additional layer of complexity compared to cluster 5, which does not display a pattern of infectious diseases. Similarly, while cluster 6 has the highest level of multimorbidity, it can be seen as less complex than clusters 1 and 5, as it lacks the complexity signature formed by the combination substance dependency-mental health conditions.

Of note, dementia, prevalent in clusters 2, 3 and 6, is a condition marked by significant cognitive and functional impairments \cite{Kalaria2016} (i.e. a high burden of illness). This condition is also recognized as one of major causes of dependency post-stroke \cite{leys2005poststroke}. Dependency, however, refers implicitly to one patient's relational network and its ability to interact with the healthcare system in order to cope with the patient workload \cite{may2014rethinking}. Situations of dependence imply therefore a shift in the relevant unit of analysis, from the patient to the group of agents, including relatives and caregivers (the patient's network), whose action tend to compensate for the different situations of dependence \cite{may2014rethinking} and prevent from complexity. More generally, by increasing capacity vis-à-vis workload, patient active or passive ability to mobilise support networks is an important aspect of patient resilience toward complexity. This ability, which depends also on patient socioeconomic environment, should be barred in mind when interpreting higher prevalence of LTCs associated with both high levels of illness and treatment such as dementia or cancer (clusters 1-3).

\subsection{ Conventional stroke risk factors, age and ethnicity (clusters 2 to 4) }

Clusters 2 to 4 form the core clusters within the typology, accounting for over half of analysed patients. Notably these clusters are characterized by a higher prevalence of conventional stroke risk factors. The opposite trends observed from clusters 2 to 4 in the median age at record of stroke and the prevalence of hypertension, diabetes, and hypercholesterolaemia, in parallel with higher representation of patients from Asian, Black, and mixed ethnicity is a marking trait of the typology (tables \ref{tab:t2} and \ref{tab:t1} figure \ref{fig:4} and supplementary figure S4). These trends highlight the importance of these conventional stroke risk factors in a lifespan, earlier records of stroke being observed in clusters characterised by higher prevalence of these conventional stroke risk factors. Importantly, the parallel increase in the proportion of patients from non-White ethnicity in these clusters not only suggests an ethnic susceptibility to these risk factors, but also an ethnic vulnerability to stoke given these risk factors \cite{howard2011traditional}. Importantly, this highlights also the opportunity for age and culturally tailored lifestyle interventions for the prevention and management of these conventional stroke risk factors \cite{wadi2022culturally}.\\

\subsection{ Low levels of multimorbidity (clusters 7 and 8)  }

Sickle cell anaemia, characterising cluster 7, is the most common and severe form of sickle cell disease and is particularly prevalent in people of African, Middle Eastern and Indian descent \cite{rees2010sickle}. Despite available strategies to prevent (and manage) stroke in sickle cell anaemia patients \cite{platt2006prevention}, its appearance as a marking trait of cluster 7 highlights its significance as a stroke risk factor in younger patients \cite{ohene1998cerebrovascular}. In a research perspective, this illustrates the ability of the employed method to disentangle longitudinal life course trajectories, enabling therefore to analyse longitudinal data in multimorbidity research \cite{shippee2012cumulative}. To this extent, it is also important to note the lower prevalence of multimorbidity (including stroke risk factors) encountered in clusters 7 and 8. This trend corresponds to the lower prevalence of multimorbidity observed in younger patients \cite{Barnett2012}, but also possibly, to a number of cryptogenic strokes \cite{hart2014embolic} as well as insufficient screening strategies of cardiovascular risk factors in this younger population \cite{lang2016impact}.


\section{Strengths and Limitations}

In order to address the limitations associated with the use of indices, recent studies on multimorbidity have adopted data-driven approaches such as cluster analysis or exploratory factor analysis \cite{ng2018patterns,sukumaran2023understanding}. While these methods offer insights into the most recurrent combinations of LTCs in analysed cohorts, their practicality remain nuanced as they provide limited evidence regarding potential clusters of patients according to the various combinations of LTCs \cite{aquino2019does}. More importantly, these methods apply to cross-sectional studies \cite{ng2018patterns} and do not allow the use longitudinal data.

The present study differs from previous published studies \cite{ng2018patterns} in both the nature of analysed data, and its methodological framework. We used longitudinal primary care EHRs representing patients' history and adopted a patient-oriented unsupervised learning method to obtain clusters of patients as opposed to clusters of LTCs. This procedure tend to make the most out of patients' history and operates a shift from a disease-oriented to a patient-oriented approach to multimorbidity. As a result, inferred patterns of multimorbidity are representative of specific groups of patients (clusters) sharing recurrent life course health trajectories. In turn, these clusters can be described in terms of their relative size, patterns of LTCs as well as socio-demographic characteristics and risk factors, addressing evidence-gaps in research on multimorbidity \cite{aquino2019does,safford2007patient}. For instance, this approach allows to characterise broad clusters such as cluster 1, or  backbone sequences of LTCs associated with distinct clusters, such as the sequence hypertension-diabetes-stroke shifted by 15 years in cluster 4 vs cluster 3, in distinct socio-demographic contexts. It allows also to identify smaller clusters characterised by recurrent sequences such as sickle cell anaemia-stroke in cluster 7.

Beside this, it is important to acknowledge limitations regarding generalization of our results. These limitations relate to methodological parameters, specificity of EHRs systems as well as characteristics of the analysed population. Methodological parameters include for instance the choice of a metrics for computing patients pairwise distances or the clustering algorithm itself. In the absence of a foundation in statistical theory on which optimal decisions can be made, these choices rely solely on investigators' aims and preferences \cite{ahlquist2012model}, which should be accounted for when handling results.

While EHRs are considered a reliable source of data and have been increasingly used in epidemiological research, surveillance, and public health planning \cite{hripcsak2013next,ehrenstein2017clinical}, results from EHRs studies remain subject to factors such as difference in data entry protocols, diagnostic coding accuracy, timeliness of updates as well as factors related to patients \cite{weiskopf2013methods}. This should be also considered when interpreting results.

Finally, it is important to consider that this study was conducted in a high income developed country, urban, multi-ethnic, deprived and young age population and therefore reflects the underlying patterns of multimorbidity in this specific setting, notably, earlier onset of multimorbidity and stroke \cite{Barnett2012,bray2018socioeconomic}. Our results need therefore to be interpreted with caution when extrapolated to other settings. For instance, the relative size of clusters are directly related to the structure of the underlying population in terms of age, gender and ethnicity. Likewise, the age at onset of stroke and other LTCs as well as their frequency is also known to be associated with levels of deprivation observed in considered populations \cite{Barnett2012,bray2018socioeconomic}.

\section{Conclusion}

This study meets the need for a better understanding of stroke-associated multimorbidity. Notably, the proposed typology does not rely on an index or a specific definition of multimorbidity, but provide a comprehensive description of the main patterns of multimorbidity characterising a limited number of patient clusters, in terms of age at record of stroke and other LTCs, gender, ethnicity and lifestyle associated risk factors. Importantly, defined clusters reflect marked trends in age, ethnicity and prevalence of conventional stroke risk factors and highlight the importance of mental health conditions in complex profiles of multimorbidity displayed in a significant proportion of patients. The typology reveals also specific combinations such as osteoarthritis and chronic pain in older patients, sickle cell anaemia or low levels of multimorbidity and risk factors in younger patients. These novel perspectives on stroke-associated multimorbidity address existing evidence gaps to inform more efficient and patient-oriented healthcare models.

\section{ Patients and method }
\subsection{ Patients }
\subsubsection{ Primary care data }

We utilised EHRs of 30 LTCs (including stroke) in adult patients aged over 18 and registered in 41 general practices in south London between April 2005 and April 2021. This included any of the selected LTCs recorded before the end of patients follow up. LTCs where selected as per the consensus definition for multimorbidity of the Academy of Medical Sciences \cite{acmedsci2023} and a definition of multimorbidity proposed by Hafezparast et al. \cite{hafezparast2021adapting} which aimed to define multimorbidity for urban, multi-ethnic, deprived and young age communities.

Accordingly, selected LTCs refer to either physical non-communicable diseases of long duration such as a cardiovascular disease or cancer, mental health conditions such as mood disorder or dementia, or infectious diseases such as HIV/AIDS or viral hepatitis.

Resulting LTCs are listed in table \ref{table:icd_codes} that includes, for instance, conventional stroke risk factors such as coronary heart disease, hypertension, and diabetes, but also LTCs \textit{a priori} less related or non-directly related to stroke such as cancers, chronic kidney disease (CKD), asthma, and chronic obstructive pulmonary disease (COPD). Patient EHR included the date at which any of the considered LTCs were first ever recorded. All patients with a record of stroke were deemed eligible for this study, regardless of the number of associated LTCs.

Associated socio-demographic variables were gender, age at record of stroke, ethnicity (Asian ethnicity, Black ethnicity, mixed and other ethnicity , and White ethnicity), quintile of the locally calculated index of multiple deprivation (IMD) 2019 and end of follow-up status (censoring or death).

Other variables/risk factors include smoking ever status, alcohol consumption (over 14 units of alcohol per week), substance use/dependency, chronic pain, hypercholesterolaemia (total cholesterol over 5.0 mmol/L) and morbid obesity. 

Data were provided by the Lambeth DataNet and approval for the analysis of fully anonymised data was granted by Lambeth Clinical Commissioning Group and Information Governance Steering Group.

\renewcommand{\arraystretch}{1.3} 
\begin{table}[!ht]
\centering
{\tiny
\begin{tabular}{|m{0.5cm}|m{3.5cm}|m{1cm}|m{4cm}|}
\hline
\textbf{No.} & \multicolumn{1}{m{3.5cm}|}{\textbf{Category}} & \textbf{ICD-10 code} & \textbf{Long-term condition} \\
\hline
1 & \multicolumn{1}{m{3.5cm}|}{Neoplasms} & C00-C97 & Cancer \\
\cline{1-2}\cline{3-4}
2 & \multirow{8}{3.5cm}{Diseases of the circulatory system} & I48 & Atrial Fibrillation \\
\cline{3-4}
3 &  & I50 & Heart Failure \\
\cline{3-4}
4 &  & I10 & Hypertension \\
\cline{3-4}
5 &  & I21 & Myocardial infarction \\
\cline{3-4}
6 &  & I73 & Peripheral Aterial/Vascular Disease \\
\cline{3-4}
7 &  & G45 & Transient cerebral ischemic attacks (TIA) \\
\cline{3-4}
8 &  & I60-I64 & Stroke \\
\cline{3-4}
9 &  & I20-I25 & Coronary Heart Disease \\
\cline{1-2}\cline{3-4}
10 & \multicolumn{1}{m{3.5cm}|}{Certain infectious and parasitic diseases} & B15-B19 & Viral Hepatitis (B\&C) \\
\cline{3-4}
11 &  & B20 & HIV/AIDS \\
\cline{1-2}\cline{3-4}
12 & \multicolumn{1}{m{3.5cm}|}{Diseases of the blood and blood-forming organs and certain disorders involving the immune mechanism} & D57 & Sickle Cell Anaemia \\
\cline{1-2}\cline{3-4}
13 & \multicolumn{1}{m{3.5cm}|}{Endocrine, nutritional and metabolic diseases} & E10-E14 & Diabetes \\
\cline{3-4}
\cline{1-2}\cline{3-4}
14& \multirow{4}{3.5cm}{Mental, Behavioral and Neurodevelopmental disorders} & F32 & Depression \\
\cline{3-4}
15 &  & F40-F41 & Anxiety Disorders \\
\cline{3-4}
16 &  & F20-F29 & Serious Mental Illness \\
\cline{3-4}
17 &  & F00-F03, G30 & Dementia/Alzheimer's \\
\cline{1-2}\cline{3-4}
18 & \multirow{3}{3.5cm}{Diseases of the nervous system} & G40-G41 & Epilepsy \\
\cline{3-4}
19 &  & G20 & Parkinson's disease \\
\cline{3-4}
20 &  & G35 & Multiple Sclerosis \\
\cline{1-2}\cline{3-4}
21 & \multirow{2}{3.5cm}{Diseases of the respiratory system} & J45-J46 & Asthma \\
\cline{3-4}
22 &  & J44 & Chronic Obstructive Pulmonary Disease (COPD) \\
\cline{1-2}\cline{3-4}
23 & \multirow{2}{3.5cm}{Diseases of the digestive system} & K50-K52 & Inflammatory Bowel Disease \\
\cline{3-4}
24 &  & K70-K76 & Liver Disease \\
\cline{1-2}\cline{3-4}
25 & \multicolumn{1}{m{3.5cm}|}{Diseases of the genitourinary system} & N18 & Chronic kidney disease stage 3 to stage 5 (CKD 3-5) \\
\cline{1-2}\cline{3-4}
26 & \multirow{3}{3.5cm}{Diseases of the musculoskeletal system and connective tissue} & M05-M06 & Rheumatoid Arthritis \\
\cline{3-4}
27 &  & M15-M19 & Osteoarthritis \\
\cline{3-4}
28 &  & M80-M82 & Osteoporosis  \\
\cline{1-2}\cline{3-4}
29 & \multicolumn{1}{m{3.5cm}|}{Diseases of the musculoskeletal system and connective tissue} & M32 & Lupus \\
\cline{1-2}\cline{3-4}
30 & \multicolumn{1}{m{3.5cm}|}{Symptoms, signs and abnormal clinical and laboratory findings, not elsewhere classified} & R41.83 & Cognitive and learning disability \\
\hline
\end{tabular}
}
\caption{Long term conditions analysed: long-term conditions are classified according to the 10\textsuperscript{th} international classification of diseases  (ICD-10)}
\label{table:icd_codes}
\end{table}
\renewcommand{\arraystretch}{1}

\subsection{ Method }

\subsubsection{ Clustering analysis }

Individual patients' records were analysed using a patient-oriented clustering approach based on the Ward's minimum variance criterion. This involves the three following steps \cite{kaufman2009finding,milligan1985examination}: i) computation of pairwise distances between patients, ii) computation of the hierarchical clustering, and iii) determination of the size of the typology.

In order to accommodate the use of longitudinal and censored patient histories, individual patient records were translated into state matrices which compile patients state indicators across analyzed LTCs. For instance, if a patient is diagnosed with hypertension at age 60, its state indicator for hypertension would be 0 from age 0 to 59, change to 1 from age 60 until the patient's death or censoring, and remain undefined thereafter. One patient state matrix is formed with as many rows as the number of analyzed LTCs ranging from age 0 to the most advance age at end of follow up in the cohort. Simplified examples of states matrices are given in supplementary figure S2.

Given individual patients state matrices, patients' pairwise dissimilarities were computed elementwise using the Jaccard distance \cite{jaccard1901etude}.
The Jaccard distances, also referred to as the binary metric, is a measure of dissimilarity between binary outcomes, such as state indicators. It is defined as one minus the Jaccard index, which is conversely a measure of similarity between sets. The Jaccard index between two indicators takes values between 0 and 1 and is computed as the number of positive matching unit of time over the number of unit either indicator has been positive.

After computation of the Ward’s hierarchical clustering, the point-biserial correlation was used to determine the optimal size of the typology within a convenient and workable range (from two to 20 clusters) \cite{milligan1985examination}. Briefly, for a given partition size, this indicator reflects the correlation between the original distance matrix (which is actually used to compute the hierarchical clustering), and a twin binary matrix in which entries are set to zero for any pair of patients belonging the the same cluster, and one otherwise. Therefore, the higher the  point-biserial correlation, the closer the actual distance matrix to the tested partition, and the better the tested partition reflects the underlying structure of the data as per this criterion. A local maximum of this coefficient across the range of tested partition size indicates an optimal partition size.

\subsubsection{ Remarks }

We used the Ward’s hierarchical agglomerative clustering method associated with the Jaccard dissimilarity index. The Ward’s method minimize the within-cluster variance, and aim at providing compact and well-defined clusters \cite{murtagh2014ward}. The Jaccard index, computed from patients state matrices (supplementary figure S2), allow to capture pairwise similarities in patients health trajectories while providing a meaningful epidemiological interpretation: it represent for two patients and a given LTC, the number of years since both patients have had a record of the investigated LTC over the number of years since either patient have had a record of this LTC until either patient end of follow-up. Therefore, the longer two patients had remained in the same state of health (given their respective age), the smaller the Jaccard distance and the more likely these patients are to be co-clustered, which is the expected result. Of note, the Jaccard distance is computed on patient pairwise common follow up. Right censoring is therefore equivalent to negative matches, which are not accounted for when computing the Jaccard distance, and is seamlessly handled when computing pairwise distances.

\subsubsection{ Statistical analysis }

LTCs, socio-demographic variables and other variables, were displayed as frequencies and percentage or median and interquartile range as appropriate. Associations between variables and clusters were tested using the Fisher exact test for categorical variables or the Kruskal–Wallis test for numeric variables (tables \ref{tab:t1} and \ref{tab:t2}).
Multivariate associations between clusters and socio-demographic variables, conventional stroke risk factors, and LTCs were estimated using logistic regressions where cluster indicators were explained by tested variables (figure \ref{fig:3}).
All computations were performed using the R language and environment for statistical computing (version 4.3.0 (2023-04-21))\cite{R}.

\section*{Author contribution}
AD and MD designed the study (conceptualization and methodology), AD, MA, XS and MD conducted investigations, XS, AL and MD carried out data curation, MD provided computer code and performed formal analysis including data visualisation and validation, MD drafted the manuscript, and AD supervised the research. MA, AD, VC, XS and CW reviewed and edited the manuscript.

\section*{Funding}

This project is funded by King’s Health Partners / Guy’s and St Thomas Charity ‘MLTC
Challenge Fund’ (grant number EIC180702) and support from the National Institute for Health and Care
Research (NIHR) under its Programme Grants for Applied Research (NIHR202339)

\section*{Data availability}

Restrictions apply to the availability of the data that support the findings of this study, since they were used under an IRB approval, and hence not publicly available.


\newgeometry{top=0cm, bottom=0cm, left=2cm , right=0cm}

\begin{landscape}\begingroup\fontsize{6}{8}\selectfont

\begin{longtable}[t]{llllllllllll}
\caption{\label{tab:t2}Distribution of socio-demographics and risk factors associated to stroke accross the clusters}\\
\toprule
Cluster \# &  & 1 & 2 & 3 & 4 & 5 & 6 & 7 & 8 & Total & P\\
\midrule
\endfirsthead
\caption[]{Distribution of socio-demographics and risk factors associated to stroke accross the clusters \textit{(continued)}}\\
\toprule
Cluster \# &  & 1 & 2 & 3 & 4 & 5 & 6 & 7 & 8 & Total & P\\
\midrule
\endhead
\midrule
\multicolumn{12}{r@{}}{\textit{(Continued on Next Page...)}}\
\endfoot
\bottomrule
\multicolumn{12}{l}{\rule{0pt}{1em}\textsuperscript{a} IMD\textsuperscript{a} quitile 1 (most deprived)-2 vs 3-5 (less deprived)}\\
\multicolumn{12}{l}{\rule{0pt}{1em}\textsuperscript{b} Long-term condition}\\
\endlastfoot
\cellcolor{gray!6}{Gender} & \cellcolor{gray!6}{Female} & \cellcolor{gray!6}{1025 (42.9)} & \cellcolor{gray!6}{969 (53.4)} & \cellcolor{gray!6}{729 (40.9)} & \cellcolor{gray!6}{581 (41.2)} & \cellcolor{gray!6}{463 (54.8)} & \cellcolor{gray!6}{404 (59.0)} & \cellcolor{gray!6}{271 (48.1)} & \cellcolor{gray!6}{133 (37.3)} & \cellcolor{gray!6}{4575 (46.5)} & \cellcolor{gray!6}{<0.001}\\
Age at onset & Median (IQR) & 59.2 (52.6 to 68.0) & 81.2 (76.1 to 86.3) & 70.7 (65.8 to 77.0) & 55.0 (49.0 to 63.0) & 47.0 (36.0 to 56.9) & 78.1 (69.5 to 85.4) & 26.8 (19.2 to 32.0) & 44.4 (41.1 to 47.0) & 65.0 (51.5 to 77.0) & <0.001\\
\cellcolor{gray!6}{Ethnicity} & \cellcolor{gray!6}{White} & \cellcolor{gray!6}{1284 (53.7)} & \cellcolor{gray!6}{1088 (60.0)} & \cellcolor{gray!6}{830 (46.6)} & \cellcolor{gray!6}{441 (31.3)} & \cellcolor{gray!6}{497 (58.8)} & \cellcolor{gray!6}{361 (52.7)} & \cellcolor{gray!6}{307 (54.5)} & \cellcolor{gray!6}{191 (53.5)} & \cellcolor{gray!6}{4999 (50.8)} & \cellcolor{gray!6}{<0.001}\\
 & Asian & 147 (6.1) & 103 (5.7) & 138 (7.7) & 131 (9.3) & 42 (5.0) & 43 (6.3) & 21 (3.7) & 20 (5.6) & 645 (6.6) & \\
\cellcolor{gray!6}{} & \cellcolor{gray!6}{Black} & \cellcolor{gray!6}{550 (23.0)} & \cellcolor{gray!6}{268 (14.8)} & \cellcolor{gray!6}{485 (27.2)} & \cellcolor{gray!6}{619 (43.9)} & \cellcolor{gray!6}{185 (21.9)} & \cellcolor{gray!6}{168 (24.5)} & \cellcolor{gray!6}{128 (22.7)} & \cellcolor{gray!6}{92 (25.8)} & \cellcolor{gray!6}{2495 (25.3)} & \cellcolor{gray!6}{}\\
 & Mixed & 78 (3.3) & 33 (1.8) & 54 (3.0) & 52 (3.7) & 32 (3.8) & 20 (2.9) & 30 (5.3) & 14 (3.9) & 313 (3.2) & \\
\cellcolor{gray!6}{} & \cellcolor{gray!6}{Other/Unknown} & \cellcolor{gray!6}{333 (13.9)} & \cellcolor{gray!6}{322 (17.8)} & \cellcolor{gray!6}{275 (15.4)} & \cellcolor{gray!6}{166 (11.8)} & \cellcolor{gray!6}{89 (10.5)} & \cellcolor{gray!6}{93 (13.6)} & \cellcolor{gray!6}{77 (13.7)} & \cellcolor{gray!6}{40 (11.2)} & \cellcolor{gray!6}{1395 (14.2)} & \cellcolor{gray!6}{}\\
 & <3 - Most deprived & 1638 (68.5) & 1190 (65.6) & 1192 (66.9) & 1038 (73.7) & 568 (67.2) & 451 (65.8) & 372 (66.1) & 232 (65.0) & 6681 (67.8) & <0.001\\
\cellcolor{gray!6}{IMD quintile} & \cellcolor{gray!6}{>2 - Least deprived} & \cellcolor{gray!6}{736 (30.8)} & \cellcolor{gray!6}{611 (33.7)} & \cellcolor{gray!6}{579 (32.5)} & \cellcolor{gray!6}{366 (26.0)} & \cellcolor{gray!6}{270 (32.0)} & \cellcolor{gray!6}{228 (33.3)} & \cellcolor{gray!6}{182 (32.3)} & \cellcolor{gray!6}{116 (32.5)} & \cellcolor{gray!6}{3088 (31.4)} & \cellcolor{gray!6}{}\\
 & (Missing) & 18 (0.8) & 13 (0.7) & 11 (0.6) & 5 (0.4) & 7 (0.8) & 6 (0.9) & 9 (1.6) & 9 (2.5) & 78 (0.8) & \\
\cellcolor{gray!6}{Smoking ever} & \cellcolor{gray!6}{1} & \cellcolor{gray!6}{1610 (67.3)} & \cellcolor{gray!6}{1044 (57.6)} & \cellcolor{gray!6}{1059 (59.4)} & \cellcolor{gray!6}{850 (60.3)} & \cellcolor{gray!6}{588 (69.6)} & \cellcolor{gray!6}{378 (55.2)} & \cellcolor{gray!6}{261 (46.4)} & \cellcolor{gray!6}{210 (58.8)} & \cellcolor{gray!6}{6000 (60.9)} & \cellcolor{gray!6}{<0.001}\\
Substance dependancy & 1 & 215 (9.0) & 39 (2.1) & 43 (2.4) & 57 (4.0) & 177 (20.9) & 22 (3.2) & 38 (6.7) & 18 (5.0) & 609 (6.2) & <0.001\\
\cellcolor{gray!6}{Alcohol dependence} & \cellcolor{gray!6}{1} & \cellcolor{gray!6}{744 (31.1)} & \cellcolor{gray!6}{330 (18.2)} & \cellcolor{gray!6}{416 (23.3)} & \cellcolor{gray!6}{326 (23.1)} & \cellcolor{gray!6}{296 (35.0)} & \cellcolor{gray!6}{151 (22.0)} & \cellcolor{gray!6}{94 (16.7)} & \cellcolor{gray!6}{88 (24.6)} & \cellcolor{gray!6}{2445 (24.8)} & \cellcolor{gray!6}{<0.001}\\
Chronic pain & 1 & 1440 (60.2) & 1060 (58.4) & 988 (55.4) & 810 (57.5) & 521 (61.7) & 589 (86.0) & 104 (18.5) & 90 (25.2) & 5602 (56.9) & <0.001\\
\cellcolor{gray!6}{Hypercholesterolemia} & \cellcolor{gray!6}{1} & \cellcolor{gray!6}{1738 (72.7)} & \cellcolor{gray!6}{1248 (68.8)} & \cellcolor{gray!6}{1342 (75.3)} & \cellcolor{gray!6}{1100 (78.1)} & \cellcolor{gray!6}{537 (63.6)} & \cellcolor{gray!6}{517 (75.5)} & \cellcolor{gray!6}{136 (24.2)} & \cellcolor{gray!6}{195 (54.6)} & \cellcolor{gray!6}{6813 (69.2)} & \cellcolor{gray!6}{<0.001}\\
Morbid obesity & 1 & 156 (6.5) & 51 (2.8) & 90 (5.1) & 177 (12.6) & 94 (11.1) & 46 (6.7) & 11 (2.0) & 12 (3.4) & 637 (6.5) & <0.001\\
\cellcolor{gray!6}{No. LTC\textsuperscript{b}} & \cellcolor{gray!6}{Median (IQR)} & \cellcolor{gray!6}{4.0 (2.0 to 5.0)} & \cellcolor{gray!6}{3.0 (2.0 to 5.0)} & \cellcolor{gray!6}{3.0 (2.0 to 5.0)} & \cellcolor{gray!6}{3.0 (2.0 to 5.0)} & \cellcolor{gray!6}{3.0 (2.0 to 5.0)} & \cellcolor{gray!6}{4.0 (3.0 to 5.0)} & \cellcolor{gray!6}{0.0 (0.0 to 1.0)} & \cellcolor{gray!6}{0.0 (0.0 to 1.0)} & \cellcolor{gray!6}{3.0 (2.0 to 5.0)} & \cellcolor{gray!6}{<0.001}\\
Total N (\%) &  & 2392 (24.3) & 1814 (18.4) & 1782 (18.1) & 1409 (14.3) & 845 (8.6) & 685 (7.0) & 563 (5.7) & 357 (3.6) & 9847 & \\*
\end{longtable}
\endgroup{}
\end{landscape}

\break
\restoregeometry
\begin{landscape}\begingroup\fontsize{8}{10}\selectfont

\begin{longtable}[t]{lllllllllll}
\caption{\label{tab:t1}Distribution of the 30 analysed long-term conditions across defined clusters}\\
\toprule
Cluster \# & 1 & 2 & 3 & 4 & 5 & 6 & 7 & 8 & Total & \\
\midrule
\endfirsthead
\caption[]{Distribution of the 30 analysed long-term conditions across defined clusters \textit{(continued)}}\\
\toprule
Cluster \# & 1 & 2 & 3 & 4 & 5 & 6 & 7 & 8 & Total & \\
\midrule
\endhead
\midrule
\multicolumn{11}{r@{}}{\textit{(Continued on Next Page...)}}\
\endfoot
\bottomrule
\multicolumn{11}{l}{\rule{0pt}{1em}\textsuperscript{a} Chronic Obstructive Pulmonary Disease}\\
\multicolumn{11}{l}{\rule{0pt}{1em}\textsuperscript{b} Human Immunodeficiency Virus/Acquired Immunodeficiency Syndrome}\\
\multicolumn{11}{l}{\rule{0pt}{1em}\textsuperscript{c} Chronic Kidney Disease stage 3 to stage 5}\\
\multicolumn{11}{l}{\rule{0pt}{1em}\textsuperscript{d} Transient Ischemic Attack}\\
\endlastfoot
\cellcolor{gray!6}{Coronary Heart Disease} & \cellcolor{gray!6}{419 (17.5)} & \cellcolor{gray!6}{335 (18.5)} & \cellcolor{gray!6}{382 (21.4)} & \cellcolor{gray!6}{267 (18.9)} & \cellcolor{gray!6}{84 (9.9)} & \cellcolor{gray!6}{140 (20.4)} & \cellcolor{gray!6}{10 (1.8)} & \cellcolor{gray!6}{19 (5.3)} & \cellcolor{gray!6}{1656 (16.8)} & \cellcolor{gray!6}{<0.001}\\
Parkinson's disease & 61 (2.6) & 58 (3.2) & 39 (2.2) & 16 (1.1) & 5 (0.6) & 17 (2.5) & 0 (0.0) & 2 (0.6) & 198 (2.0) & <0.001\\
\cellcolor{gray!6}{Heart Failure} & \cellcolor{gray!6}{296 (12.4)} & \cellcolor{gray!6}{264 (14.6)} & \cellcolor{gray!6}{278 (15.6)} & \cellcolor{gray!6}{190 (13.5)} & \cellcolor{gray!6}{63 (7.5)} & \cellcolor{gray!6}{113 (16.5)} & \cellcolor{gray!6}{12 (2.1)} & \cellcolor{gray!6}{7 (2.0)} & \cellcolor{gray!6}{1223 (12.4)} & \cellcolor{gray!6}{<0.001}\\
Osteoarthritis & 566 (23.7) & 401 (22.1) & 421 (23.6) & 324 (23.0) & 138 (16.3) & 683 (99.7) & 26 (4.6) & 24 (6.7) & 2583 (26.2) & <0.001\\
\cellcolor{gray!6}{Asthma} & \cellcolor{gray!6}{441 (18.4)} & \cellcolor{gray!6}{244 (13.5)} & \cellcolor{gray!6}{79 (4.4)} & \cellcolor{gray!6}{72 (5.1)} & \cellcolor{gray!6}{342 (40.5)} & \cellcolor{gray!6}{38 (5.5)} & \cellcolor{gray!6}{27 (4.8)} & \cellcolor{gray!6}{12 (3.4)} & \cellcolor{gray!6}{1255 (12.7)} & \cellcolor{gray!6}{<0.001}\\
Serious Mental Illness & 280 (11.7) & 50 (2.8) & 26 (1.5) & 32 (2.3) & 46 (5.4) & 2 (0.3) & 16 (2.8) & 7 (2.0) & 459 (4.7) & <0.001\\
\cellcolor{gray!6}{Hypertension} & \cellcolor{gray!6}{1430 (59.8)} & \cellcolor{gray!6}{1206 (66.5)} & \cellcolor{gray!6}{1424 (79.9)} & \cellcolor{gray!6}{1301 (92.3)} & \cellcolor{gray!6}{352 (41.7)} & \cellcolor{gray!6}{485 (70.8)} & \cellcolor{gray!6}{64 (11.4)} & \cellcolor{gray!6}{82 (23.0)} & \cellcolor{gray!6}{6344 (64.4)} & \cellcolor{gray!6}{<0.001}\\
Depression & 844 (35.3) & 317 (17.5) & 200 (11.2) & 274 (19.4) & 531 (62.8) & 84 (12.3) & 64 (11.4) & 26 (7.3) & 2340 (23.8) & <0.001\\
\cellcolor{gray!6}{Anxiety Disorders} & \cellcolor{gray!6}{517 (21.6)} & \cellcolor{gray!6}{329 (18.1)} & \cellcolor{gray!6}{146 (8.2)} & \cellcolor{gray!6}{233 (16.5)} & \cellcolor{gray!6}{570 (67.5)} & \cellcolor{gray!6}{91 (13.3)} & \cellcolor{gray!6}{79 (14.0)} & \cellcolor{gray!6}{23 (6.4)} & \cellcolor{gray!6}{1988 (20.2)} & \cellcolor{gray!6}{<0.001}\\
Cancer & 460 (19.2) & 387 (21.3) & 351 (19.7) & 133 (9.4) & 85 (10.1) & 88 (12.8) & 9 (1.6) & 25 (7.0) & 1538 (15.6) & <0.001\\
\cellcolor{gray!6}{Peripheral Aterial/Vascular Disease} & \cellcolor{gray!6}{169 (7.1)} & \cellcolor{gray!6}{111 (6.1)} & \cellcolor{gray!6}{181 (10.2)} & \cellcolor{gray!6}{113 (8.0)} & \cellcolor{gray!6}{25 (3.0)} & \cellcolor{gray!6}{52 (7.6)} & \cellcolor{gray!6}{1 (0.2)} & \cellcolor{gray!6}{1 (0.3)} & \cellcolor{gray!6}{653 (6.6)} & \cellcolor{gray!6}{<0.001}\\
Diabetes & 626 (26.2) & 409 (22.5) & 710 (39.8) & 831 (59.0) & 177 (20.9) & 190 (27.7) & 26 (4.6) & 37 (10.4) & 3006 (30.5) & <0.001\\
\cellcolor{gray!6}{CKD\textsuperscript{a} Stage3-5} & \cellcolor{gray!6}{444 (18.6)} & \cellcolor{gray!6}{514 (28.3)} & \cellcolor{gray!6}{565 (31.7)} & \cellcolor{gray!6}{423 (30.0)} & \cellcolor{gray!6}{107 (12.7)} & \cellcolor{gray!6}{207 (30.2)} & \cellcolor{gray!6}{22 (3.9)} & \cellcolor{gray!6}{14 (3.9)} & \cellcolor{gray!6}{2296 (23.3)} & \cellcolor{gray!6}{<0.001}\\
Viral Hepatitis (B\&C) & 105 (4.4) & 10 (0.6) & 9 (0.5) & 21 (1.5) & 23 (2.7) & 6 (0.9) & 6 (1.1) & 5 (1.4) & 185 (1.9) & <0.001\\
\cellcolor{gray!6}{Inflammatory Bowel Disease} & \cellcolor{gray!6}{58 (2.4)} & \cellcolor{gray!6}{23 (1.3)} & \cellcolor{gray!6}{16 (0.9)} & \cellcolor{gray!6}{18 (1.3)} & \cellcolor{gray!6}{19 (2.2)} & \cellcolor{gray!6}{14 (2.0)} & \cellcolor{gray!6}{4 (0.7)} & \cellcolor{gray!6}{1 (0.3)} & \cellcolor{gray!6}{153 (1.6)} & \cellcolor{gray!6}{<0.001}\\
Liver Disease & 99 (4.1) & 17 (0.9) & 14 (0.8) & 22 (1.6) & 37 (4.4) & 6 (0.9) & 3 (0.5) & 1 (0.3) & 199 (2.0) & <0.001\\
\cellcolor{gray!6}{Epilepsy} & \cellcolor{gray!6}{378 (15.8)} & \cellcolor{gray!6}{58 (3.2)} & \cellcolor{gray!6}{61 (3.4)} & \cellcolor{gray!6}{66 (4.7)} & \cellcolor{gray!6}{71 (8.4)} & \cellcolor{gray!6}{38 (5.5)} & \cellcolor{gray!6}{46 (8.2)} & \cellcolor{gray!6}{14 (3.9)} & \cellcolor{gray!6}{732 (7.4)} & \cellcolor{gray!6}{<0.001}\\
Osteopotosis & 94 (3.9) & 206 (11.4) & 118 (6.6) & 29 (2.1) & 23 (2.7) & 67 (9.8) & 2 (0.4) & 5 (1.4) & 544 (5.5) & <0.001\\
\cellcolor{gray!6}{Dementia/Alzheimer's} & \cellcolor{gray!6}{277 (11.6)} & \cellcolor{gray!6}{460 (25.4)} & \cellcolor{gray!6}{352 (19.8)} & \cellcolor{gray!6}{132 (9.4)} & \cellcolor{gray!6}{42 (5.0)} & \cellcolor{gray!6}{172 (25.1)} & \cellcolor{gray!6}{4 (0.7)} & \cellcolor{gray!6}{12 (3.4)} & \cellcolor{gray!6}{1451 (14.7)} & \cellcolor{gray!6}{<0.001}\\
Sickle Cell Anaemia & 11 (0.5) & 1 (0.1) & 1 (0.1) & 3 (0.2) & 2 (0.2) & 1 (0.1) & 41 (7.3) & 0 (0.0) & 60 (0.6) & <0.001\\
\cellcolor{gray!6}{Multiple Sclerosis} & \cellcolor{gray!6}{12 (0.5)} & \cellcolor{gray!6}{0 (0.0)} & \cellcolor{gray!6}{1 (0.1)} & \cellcolor{gray!6}{0 (0.0)} & \cellcolor{gray!6}{7 (0.8)} & \cellcolor{gray!6}{0 (0.0)} & \cellcolor{gray!6}{2 (0.4)} & \cellcolor{gray!6}{0 (0.0)} & \cellcolor{gray!6}{22 (0.2)} & \cellcolor{gray!6}{<0.001}\\
HIV/AIDS\textsuperscript{b} & 85 (3.6) & 1 (0.1) & 3 (0.2) & 10 (0.7) & 11 (1.3) & 0 (0.0) & 2 (0.4) & 1 (0.3) & 113 (1.1) & <0.001\\
\cellcolor{gray!6}{Myocardial infarction} & \cellcolor{gray!6}{222 (9.3)} & \cellcolor{gray!6}{163 (9.0)} & \cellcolor{gray!6}{238 (13.4)} & \cellcolor{gray!6}{127 (9.0)} & \cellcolor{gray!6}{48 (5.7)} & \cellcolor{gray!6}{66 (9.6)} & \cellcolor{gray!6}{8 (1.4)} & \cellcolor{gray!6}{7 (2.0)} & \cellcolor{gray!6}{879 (8.9)} & \cellcolor{gray!6}{<0.001}\\
Atrial Fibrillation & 365 (15.3) & 491 (27.1) & 388 (21.8) & 141 (10.0) & 74 (8.8) & 184 (26.9) & 13 (2.3) & 16 (4.5) & 1672 (17.0) & <0.001\\
\cellcolor{gray!6}{Lupus} & \cellcolor{gray!6}{34 (1.4)} & \cellcolor{gray!6}{1 (0.1)} & \cellcolor{gray!6}{4 (0.2)} & \cellcolor{gray!6}{4 (0.3)} & \cellcolor{gray!6}{7 (0.8)} & \cellcolor{gray!6}{1 (0.1)} & \cellcolor{gray!6}{1 (0.2)} & \cellcolor{gray!6}{0 (0.0)} & \cellcolor{gray!6}{52 (0.5)} & \cellcolor{gray!6}{<0.001}\\
COPD\textsuperscript{c} & 337 (14.1) & 249 (13.7) & 152 (8.5) & 60 (4.3) & 85 (10.1) & 54 (7.9) & 9 (1.6) & 8 (2.2) & 954 (9.7) & <0.001\\
\cellcolor{gray!6}{Rheumatoid Arthritis} & \cellcolor{gray!6}{89 (3.7)} & \cellcolor{gray!6}{29 (1.6)} & \cellcolor{gray!6}{19 (1.1)} & \cellcolor{gray!6}{29 (2.1)} & \cellcolor{gray!6}{15 (1.8)} & \cellcolor{gray!6}{21 (3.1)} & \cellcolor{gray!6}{2 (0.4)} & \cellcolor{gray!6}{1 (0.3)} & \cellcolor{gray!6}{205 (2.1)} & \cellcolor{gray!6}{<0.001}\\
TIA\textsuperscript{d} & 443 (18.5) & 283 (15.6) & 425 (23.8) & 165 (11.7) & 109 (12.9) & 111 (16.2) & 12 (2.1) & 9 (2.5) & 1557 (15.8) & <0.001\\
\cellcolor{gray!6}{Cognitive and learning disability} & \cellcolor{gray!6}{49 (2.0)} & \cellcolor{gray!6}{2 (0.1)} & \cellcolor{gray!6}{3 (0.2)} & \cellcolor{gray!6}{9 (0.6)} & \cellcolor{gray!6}{9 (1.1)} & \cellcolor{gray!6}{2 (0.3)} & \cellcolor{gray!6}{11 (2.0)} & \cellcolor{gray!6}{0 (0.0)} & \cellcolor{gray!6}{85 (0.9)} & \cellcolor{gray!6}{<0.001}\\
Total N (\%) & 2392 (24.3) & 1814 (18.4) & 1782 (18.1) & 1409 (14.3) & 845 (8.6) & 685 (7.0) & 563 (5.7) & 357 (3.6) & 9847 & \\*
\end{longtable}
\endgroup{}
\end{landscape}
\break
\restoregeometry


\setstretch{0.9}
\bibliographystyle{unsrt}


\end{document}